\documentclass[aps,prb,twocolumn,superscriptaddress]{revtex4}
\usepackage{amsfonts}
\usepackage{graphicx}
\usepackage{epstopdf}
\usepackage{epsfig}
\usepackage{amsmath}
\usepackage{color}
\usepackage{makecell}
\usepackage[normalem]{ulem}

\begin{document}

\title{Non-equilibrium transport in the pseudospin-1 Dirac-Weyl system}

\author{Cheng-Zhen Wang}
\affiliation{School of Electrical, Computer and Energy Engineering, Arizona State University, Tempe, Arizona 85287, USA}

\author{Hong-Ya Xu}
\affiliation{School of Electrical, Computer and Energy Engineering, Arizona State University, Tempe, Arizona 85287, USA}

\author{Liang Huang}
\affiliation{School of Physical Science and Technology, and Key Laboratory for Magnetism and Magnetic Materials of MOE, Lanzhou University, Lanzhou, Gansu 730000, China}

\author{Ying-Cheng Lai} \email{Ying-Cheng.Lai@asu.edu}
\affiliation{School of Electrical, Computer and Energy Engineering, Arizona State University, Tempe, Arizona 85287, USA}
\affiliation{Department of Physics, Arizona State University, Tempe, Arizona 85287, USA}

\date{\today}
\begin{abstract}

Solid state materials hosting pseudospin-1 quasiparticles have attracted a great deal of recent attention. In these materials, the energy band contains of a pair of Dirac cones and a flat band through the connecting point of the cones. As the ``caging'' of carriers with a zero group velocity, the flat band itself has zero conductivity. However, in a non-equilibrium situation where a constant electric field is suddenly switched on, the flat band can enhance the resulting current in both the linear and nonlinear response regimes through distinct physical mechanisms. Using the ($2+1$) dimensional pseudospin-$1$ Dirac-Weyl system as a concrete setting, we demonstrate that, in the weak field regime, the interband current is about twice larger than that for pseudospin-1/2 system due to the interplay between the flat band and the negative band, with the scaling behavior determined by the Kubo formula. In the strong field regime, the intraband current is $\sqrt{2}$ times larger than that in the pseudospin-1/2 system, due to the additional contribution from particles residing in the flat band. In this case, the current and field follows the scaling law associated with Landau-Zener tunneling. These results provide a better understanding of the role of the flat band in non-equilibrium transport and are experimentally testable using electronic or photonic systems.

\end{abstract}


\maketitle

\section{Introduction} \label{sec:intro}

Solid state materials, due to the rich variety of their lattice structures and
intrinsic symmetries~\cite{bradlyn2016beyond,beenakker2016bringing}, can
accommodate quasiparticles that lead to quite unconventional and interesting
physical phenomena. The materials and the resulting exotic quasiparticles
constitute the so-called ``material universe.'' Such materials range from
graphene that hosts Dirac fermions~\cite{neto2009electronic} to 3D topological
insulators~\cite{hasan2010colloquium,qi2011topological} and 3D Dirac and
Weyl semimetals~\cite{xu2015discovery,lv2015experimental}, in which
the quasiparticles are relativistic pseudospin-$1/2$ fermions. Recently,
Dirac-like pseudospin-1 particles have attracted much attention~\cite{Bercioux2009,shen2010single,urban2011barrier,dora2011lattice,
goldman2011topological,guzman2014experimental,Li2014,Giovannetti2015,
vicencio2015observation,mukherjee2015observation,taie2015coherent,diebel2016conical,paavilainen2016coexisting,
zhu2016blue,fang2016klein,Malcolm2016,xu2016,Tsuch2016,XL2017,Fang2017}, which
are associated with a unique type of energy band structure: a pair of
Dirac cones with a flat band through the conical connecting point.
Materials that can host pseudospin-1 particles include particularly
engineered photonic crystals~\cite{fang2016klein,guzman2014experimental,
vicencio2015observation,mukherjee2015observation,diebel2016conical},
optical dice or Lieb lattices with loaded ultracold
atoms~\cite{Bercioux2009,shen2010single,urban2011barrier,goldman2011topological,Raoux2014}, and
certain electronic materials~\cite{Li2014,Giovannetti2015,paavilainen2016coexisting,zhu2016blue}.
In contrast to the Dirac cone system with massless pseudospin-$1/2$ particles
that exhibit conventional relativistic quantum phenomena, in pseudospin-$1$
systems an array of quite unusual physical phenomena can arise, such as
super-Klein tunneling associated with one-dimensional barrier
transmission~\cite{shen2010single,dora2011lattice,fang2016klein},
diffraction-free wave propagation and novel conical
diffraction~\cite{guzman2014experimental,mukherjee2015observation,
vicencio2015observation,diebel2016conical}, unconventional Anderson
localization~\cite{chalker2010anderson,bodyfelt2014flatbands,Fang2017},
flat-band ferromagnetism~\cite{taie2015coherent}, unconventional
Landau-Zener Bloch oscillations~\cite{KF2016}, and
peculiar topological phases under external gauge fields or spin-orbit
coupling~\cite{goldman2011topological,wang2011nearly,aoki1996hofstadter,
weeks2010topological}. The aim of this paper is to present the phenomenon
of enhanced non-equilibrium quantum transport of pseudospin-1 particles.

Quantum transport beyond the linear response and equilibrium regime is
of great practical importance, especially in device research and development.
There have been works on nonlinear and non-equilibrium transport of
relativistic pseudospin-$1/2$ particles in Dirac and Weyl materials. For
example, when graphene is subject to a constant electric field, the dynamical
evolution of the current after the field is turned on exhibits a remarkable
minimal conductivity behavior~\cite{lewkowicz2009dynamics}. The scaling
behavior of nonlinear electric transport in graphene due to the dynamical
Landau-Zener tunneling or the Schwinger pair creation mechanism has also been
investigated~\cite{rosenstein2010ballistic,dora2010nonlinear}. Under a
strong electrical field, due to the Landau-Zener transition, a topological
insulator or graphene can exhibit a quantization breakdown phenomenon in
the spin Hall conductivity~\cite{dora2011dynamics}. More recently,
non-equilibrium electric transport beyond the linear response regime in
3D Weyl semimetals has been studied~\cite{vajna2015nonequilibrium}.
In these works, the quasiparticles are relativistic pseudospin-1/2
fermions arising from the Dirac or Weyl system with a conical type of
dispersion in their energy momentum spectrum.

In this paper, we study the transport dynamics of pseudospin-1
quasiparticles that arise in material systems with a pair of Dirac cones
and a flat band through their connecting point. Under the equilibrium
condition and in the absence of disorders, the flat band acts as a perfect
``caging'' of carriers with zero group velocity and hence it contributes
little to the conductivity~\cite{vigh2013diverging,hausler2015flat,
louvet2015origin}. However, as we will show in this paper, the flat band
can have a significant effect on the non-equilibrium transport dynamics.
Through numerical and analytic calculation of the current evolution for
both weak and strong electric fields, we find the general
phenomenon of current enhancement as compared with that associated with
non-equilibrium transport of pseudospin-1/2 particles. In particular,
for a weak field, the interband current is twice as large as that for
pseudospin-1/2 system due to the interference between particles from the flat
band and from the negative band, the scaling behavior of which agrees with
that determined by the Kubo formula. For a strong field, the intraband current
is $\sqrt{2}$ times larger than that in the pseudospin-1/2 system, as a result 
of the additional contribution from the particles residing in the flat band.
In this case, the physical origin of the scaling behavior of the
current-field relation can be attributed to Landau-Zener tunneling.
Our findings suggest that, in general, the conductivity of pseudospin-1
materials can be higher than that of pseudospin-$1/2$ materials in the
nonequilibrium transport regime.

\section{Pseudospin-1 Hamiltonian and current} \label{sec:Hamiltonian}

We consider a system of 2D noninteracting, Dirac-like pseudospin-1 particles
subject to a uniform, constant electric field applied in the $x$ direction.
The system is described by the generalized Dirac-Weyl
Hamiltonian~\cite{xu2016,urban2011barrier}. The electric field, switched on
at $t=0$, can be incorporated into the Hamiltonian through a time-dependent
vector potential~\cite{lewkowicz2009dynamics,rosenstein2010ballistic,dora2010nonlinear,dora2011dynamics,vajna2015nonequilibrium,cohen2008schwinger,ishikawa2010nonlinear,lee2014nonlinear}:
$\boldsymbol{A}(t)=[A(t), 0, 0]$, where $A(t)=-Et\Theta (t)$. The
resulting Hamiltonian is
\begin{equation} \label{eq:Hamiltonian}
H=v_F \{S_x [p_x - qA(t)] + S_y p_y\},
\end{equation}
where $v_F$ is the Fermi velocity of the pseudospin-1 particle from the
Dirac cones, $q=-e$ $(e>0)$ is the electronic charge,
$\boldsymbol{S}=(S_x, S_y, S_z)$ is a vector of matrices with components
\[ S_x=\frac{1}{\sqrt{2}}
\begin{bmatrix}
    0 & 1 & 0 \\
    1 & 0 & 1 \\
    0 & 1 & 0
\end{bmatrix},
S_y=\frac{1}{\sqrt{2}}
\begin{bmatrix}
    0 & -i & 0 \\
    i & 0 & -i \\
    0 & i & 0
\end{bmatrix},
\]
\text{and}
\[S_z=
\begin{bmatrix}
    1 & 0 & 0 \\
    0 & 0 & 0 \\
    0 & 0 & -1
\end{bmatrix}.
\]
The three matrices form a complete representation of pseudospin-1 particles,
which satisfy the angular momentum commutation relations
$[S_l, S_m]=i\epsilon_{lmn}S_n$ with three eigenvalues: $s=\pm1,0$, where
$\epsilon_{lmn}$ is the Levi-Civita symbol. However, the matrices do not follow
the Clifford algebra underlying spin-1/2 particles. The corresponding time
dependent wave equation is
\begin{equation} \label{eq:wave_equation}
i\hbar \partial_t \Psi_{p}(t) = H\Psi_{p}(t).
\end{equation}
Under the unitary transformation
\[U=
\begin{bmatrix}
    \frac{1}{2}e^{-i\theta} & -\frac{1}{\sqrt{2}}e^{-i\theta}  & \frac{1}{2}e^{{-i\theta}} \\
    \frac{\sqrt{2}}{2}      & 0                                & -\frac{\sqrt{2}}{2}  \\
    \frac{1}{2}e^{i\theta}  & \frac{1}{\sqrt{2}}e^{i\theta}    & \frac{1}{2}e^{{i\theta}}
\end{bmatrix}
\]
with $\tan\theta = p_y/[p_x-qA(t)]$, we can rewrite
Eq.~(\ref{eq:wave_equation}) in the basis of adiabatic energy as
\begin{align} \label{eq:Diracoriginal}
i\hbar \partial_t \Phi_p(t) = &\big[S_z \epsilon_p(t)
+ S_x \sqrt{2}C_0(t)\big]\Phi_p(t),
\end{align}
where $\Phi_p(t)=U^{\dagger}\Psi_p(t)=[\alpha_p(t),\gamma_p(t),\beta_p(t)]^T$,
$C_0(t)={\hbar v_F^{2}p_y eE}/{\sqrt{2}\epsilon_p^2(t)}$, and 
$\epsilon_p = v_F \sqrt{(p_x - eEt)^2 + p_y^2}$. Initially at
$t = 0$, the negative band is assumed to be fully filled:
$\Phi_p(t=0) = [0, 0, 1]^T$. From the equation of motion, we obtain the
current operator in the original basis as
$J_x=-e\nabla_{\boldsymbol{p}}H=-ev_F S_x$. In the transformed adiabatic
energy base, the current operator is
\begin{equation} \label{eq:J_x}
J_x=-ev_F(S_z\cos\theta - S_y \sin\theta).
\end{equation}
We thus have the current density for a certain state as
\begin{align} \label{eq:current1}
\langle J_x \rangle _p(t)&=-ev_F\big\{\cos\theta[|\alpha_p(t)|^2 - |\beta_p(t)|^2] \nonumber \\
&- \sqrt{2}\sin\theta \mbox{Re}[i\alpha_p(t)\gamma_p^{*}(t)+i\gamma_p(t)\beta_p^{*}(t)]
\big\}.
\end{align}
In Eq.~(\ref{eq:current1}), the first term is related to the particle number
distribution associated with the positive and negative bands, which is the intraband or conduction current. The second term in 
Eq.~(\ref{eq:current1}) characterizes the interference between particles 
from distinct bands, which is related to the phenomenon of relativistic 
Zitterbewegung and can be appropriately called the interband or polarization 
current.

To assess the contribution of a band (i.e., positive, flat, or
negative) to the interband current, we seek to simplify the current
expression. Through some algebraic substitutions, we get
\begin{align}
\partial_t |\alpha_p(t)|^2 = 2\mbox{Re} [\alpha_p(t) \partial_t \alpha_p^*(t)], \nonumber\\
\partial_t |\gamma_p(t)|^2 = 2\mbox{Re} [\gamma_p(t) \partial_t \gamma^*_p(t)]. \nonumber
\end{align}
From the Dirac equation (\ref{eq:Diracoriginal}), we have
\begin{align}
\hbar \alpha_p(t) \partial_t \alpha^*_p(t) = i\epsilon_p \alpha_p(t) \alpha^*_p(t) + iC_0 \alpha_p(t) \gamma^*_p(t), \nonumber\\
\hbar \gamma_p(t) \partial_t \gamma^*_p(t) = iC_0 \gamma_p(t) \alpha^*_p(t) + iC_0 \gamma_p(t) \beta^*_p(t), \nonumber
\end{align}
which gives
\begin{eqnarray}
\nonumber
\mbox{Re} [i\alpha_p(t) \gamma^*_p(t)] & = & \frac{\hbar}{2C_0}\partial_t |\alpha_p(t)|^2, \nonumber \\
\mbox{Re} [i\gamma_p(t) \beta^*_p(t)]  & = & \frac{\hbar}{2C_0}\big[ \partial_t |\alpha_p(t)|^2 + \partial_t |\gamma_p(t)|^2 \big].
\end{eqnarray}
Using the total probability conservation
$|\alpha_p|^2 + |\gamma_p|^2 + |\beta_p|^2 = 1$, we finally arrive at the
following current expression
\begin{align} \label{eq:current2}
\langle J_x \rangle _p (t)
&= -ev_F \Big\{ \frac{v_F (p_x - eEt)}{\epsilon_p(t)}\big[2|\alpha_p(t)|^2 + |\gamma_p(t)|^2 - 1\big] \nonumber \\
& - \frac{\epsilon_p(t)}{v_F e E}\big(2\partial_t |\alpha_p|^2 + \partial_t |\gamma_p|^2\big) \Big\},
\end{align}
where the third term in the first part that is independent of the particle
distribution vanishes after an integration over the momentum space.

For convenience, in our numerical calculations we use dimensionless
quantities, which we obtain by introducing the scale $\Delta$, 
the characteristic energy of the system. The dimensionless time, electric
field, momentum, energy, and coefficient are
\begin{eqnarray}
\nonumber
\tilde{t} & = & \Delta t/\hbar, \\ \nonumber
\tilde{E} & = & ev_F\hbar E/\Delta^2, \\ \nonumber
\tilde{p} & = & v_Fp/\Delta, \\ \nonumber
\tilde{\epsilon} & = & \sqrt{(\tilde{p}_x - \tilde{E}\tilde{t})^2
+ \tilde{p}_y^2}, \\ \nonumber
\tilde{C}_0 & = & \tilde{E}\tilde{p}_y/\sqrt{2}[(\tilde{p}_x -
\tilde{E}\tilde{t})^2 + \tilde{p}_y^2],
\end{eqnarray}
respectively. The dimensionless current $\tilde{J}$ can be expressed
in units of $e\Delta^2/v_F \hbar^2 \pi^2$.

\section{Weak field regime: enhancement of interband current}
\label{sec:weak_field}

\begin{figure}
\centering
\includegraphics[width=\linewidth]{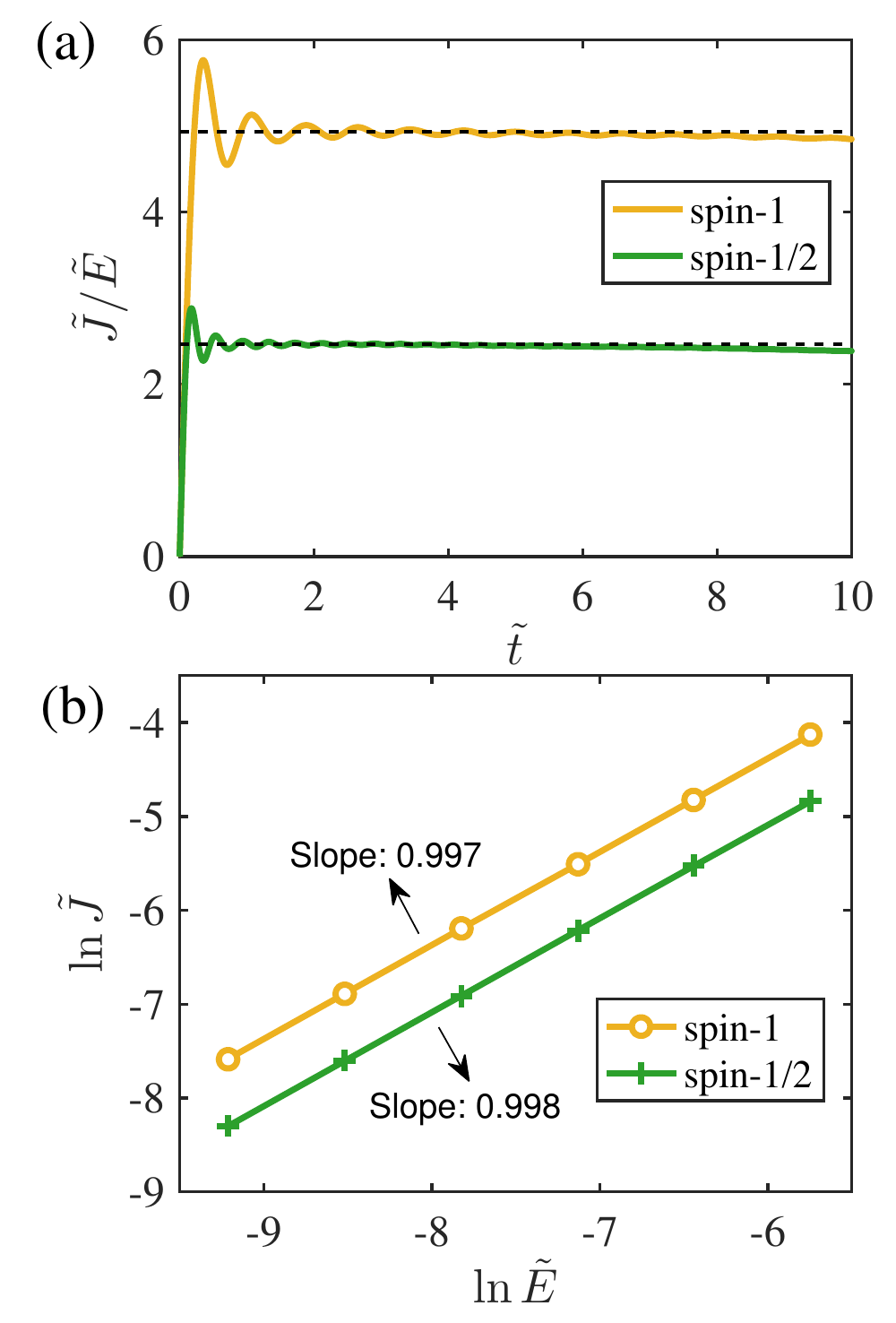}
\caption{ {\bf Interband current in pseudospin-1 and pseudospin-1/2 systems}.
(a) Evolution of the total current to electric field ratio
$\tilde{J}/\tilde{E}$ with time $\tilde{t}$ for pseudospin-1 and 1/2
systems for a fixed electric field $\tilde{E}=0.0004$, where the dashed
lines denote the theoretical values $\pi^2/2$ and $\pi^2/4$ for the
pseudospin-1 and pseudospin-1/2 systems, respectively. The yellow and green
lines represent the respective numerical results.
(b) The total current $\tilde{J}$ versus the electric field $\tilde{E}$
at time $\tilde{t}=2$ for the two systems. Comparing with the pseudospin-1/2
system, the interband current in the pseudospin-1 system is greatly enhanced.}
\label{fig:Interband_J_tE}
\end{figure}

In the weak field regime, the intraband current is negligible
as compared to the interband current due to the fewer number of conducting 
particles~\cite{rosenstein2010ballistic,dora2010nonlinear} (see Appendix B 
for an explanation and representative results). 
In particular, the interband current for a certain state can be expressed as
\begin{displaymath}
J_p^{inter} = \frac{\epsilon_p(t)}{E}[2\partial_t |\alpha_p|^2 +
\partial_t |\gamma_p|^2].
\end{displaymath}
For pseudospin-1/2 particles, the interband current has only the
first term~\cite{dora2010nonlinear}. The additional term
$[\epsilon_p(t)/E]\partial_t |\gamma_p|^2$ is unique for pseudospin-1
particles. To reveal the scaling behavior of the interband current and to
assess the role of the positive and the flat bands in the current, we impose
the weak field approximation: $|p|=\sqrt{p_x^2 + p_y^2} \gg eEt $ everywhere
except in the close vicinity of the Dirac point, which allows us to obtain
an analytic expression for the interband current. Under the approximation,
the coefficients $\epsilon_p$ and $C_0$ become $\epsilon_p\approx v_F p$ and
$C_0\approx \hbar p_y e E/(\sqrt{2}p^2)$, which are time independent.
Substituting these approximations into Eq.~(\ref{eq:Diracoriginal}), we
obtain the three components of the time dependent state $\Phi_p(t)$ as
\begin{align}
&\alpha_p(t) = \frac{1}{2} [\cos \omega t + m_0^2(\cos \omega t - 1) - 1], \\
&\beta_p(t) = \frac{1}{2} [\cos \omega t - 2m_0 \sin\omega t - m_0^2 [\cos\omega t - 1] + 1], \\
&\gamma_p(t) = \frac{1 + m_0^2}{2C_0} [-i\hbar \omega \sin \omega t - \epsilon_p(\cos \omega t - 1)].
\end{align}
The interband current contains two parts:
\begin{equation} \label{eq:interpositive}
J_p^{\alpha} = 2\frac{\epsilon_p C_0^4 \omega}{E(\epsilon_p^2 + 2C_0^2)^2}(2\sin \omega t - \sin 2\omega t),
\end{equation}
and
\begin{equation}\label{eq:interflat}
J_p^{\gamma} = 2\frac{\epsilon_p C_0^2 \omega}{E(\epsilon_p^2 + 2C_0^2)^2}(\epsilon_p^2 \sin \omega t + C_0^2 \sin 2\omega t),
\end{equation}
which correspond to contributions from the positive and the flat bands,
respectively, where $\omega = \sqrt{\epsilon_p^2 + 2C_0^2}/\hbar$. For
sufficiently weak field such that the off diagonal term is small compared
with the diagonal term in Eq.~(\ref{eq:Diracoriginal}), we have
$\epsilon_p^2 \gg 2C_0^2$, i.e.,
\begin{displaymath}
v_F^2 p^2 \gg \frac{p_y^2}{p^2}\frac{\hbar^2 e^2 E^2}{p^2}.
\end{displaymath}
In this case, the contribution from the positive band is nearly zero and the
flat band contribution is
\begin{align} \label{eq:interflat_approx}
J_p^{\gamma} \approx 2\frac{\epsilon_p^3 C_0^2 \omega}{E(\epsilon_p^2 + 2C_0^2)^2}\sin{(\omega t)} \approx  e^2 \hbar E \frac{\sin^2 \theta}{p^2}\sin{(\frac{v_Fpt}{\hbar})}.
\end{align}
The total positive band contribution over the momentum space is negligibly
small, so the flat band contributes dominantly to the total interband current:
\begin{align} \label{eq:intercurrent}
J_{inter} & = \frac{1}{\pi^2 \hbar^2}\iint e^2 \hbar E \frac{\sin^2 \theta}{p}\sin{(\frac{v_F pt}{\hbar})} d\theta dp \nonumber \\
& = \frac{e^2}{2\hbar}E  = \frac{e\Delta^2}{v_F \hbar^2 \pi^2} \cdot \frac{\pi^2}{2}\tilde{E}.
\end{align}
The dimensionless current is given by
\begin{align}
\tilde{J} = \frac{\pi^2}{2}\tilde{E}.
\end{align}

\begin{figure}
\centering
\includegraphics[width=\linewidth]{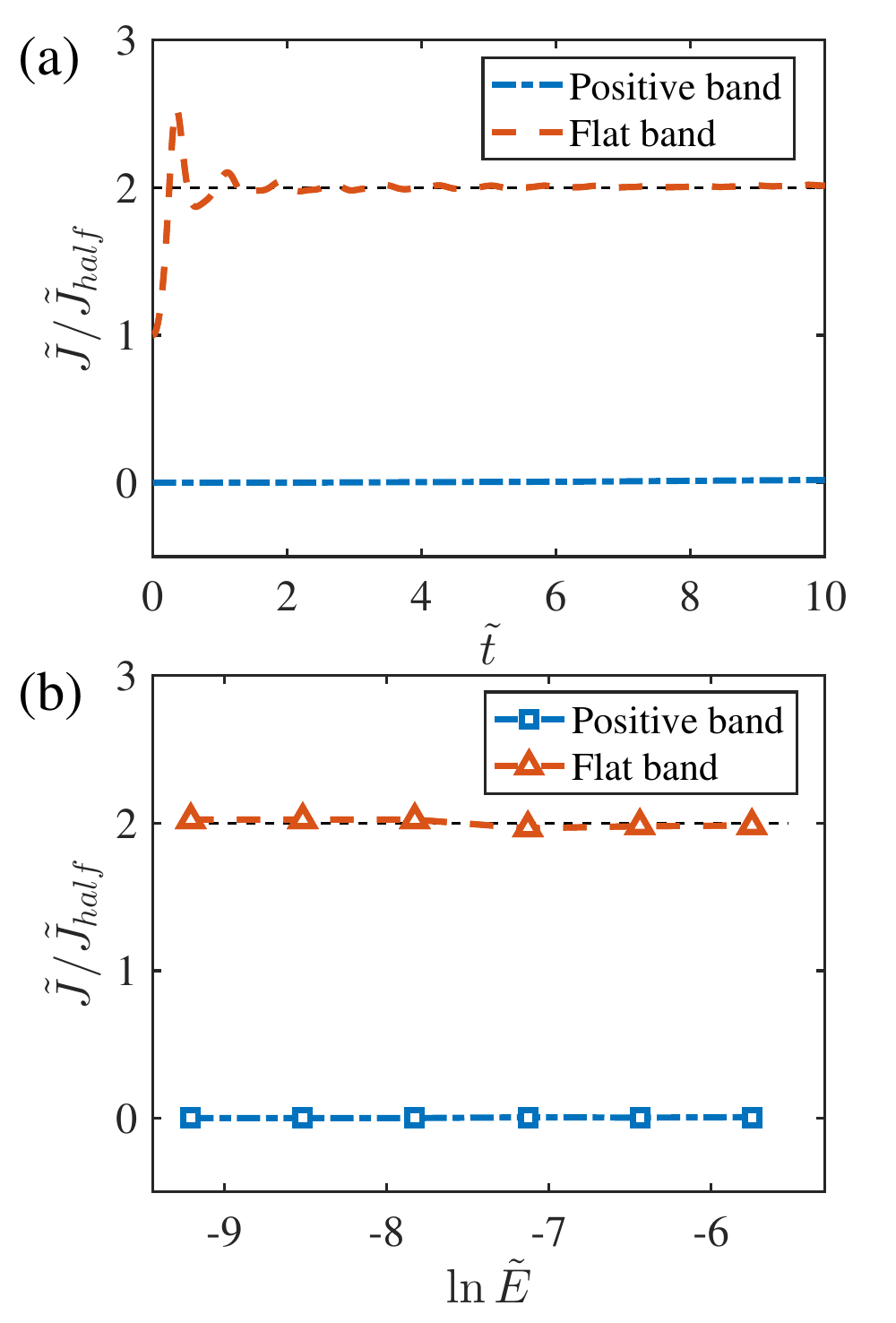}
\caption{ {\bf Origin of interband current in the pseudospin-1 system.}
(a) Ratio between interband currents from the pseudospin-1 and
pseudospin-1/2 systems as a function of time for electric field
strength $\tilde{E}=0.0004$, (b) current ratio versus $\tilde{E}$ for
fixed time $\tilde{t}=2$. The black dashed lines are theoretical results,
and the red and blue lines are for flat and positive bands, respectively.
These results indicate that, for the pseudospin-1 system, the flat band
is the sole contributor to the interband current.}
\label{fig:Interband_flat}
\end{figure}

To verify the analytical prediction Eq.~(\ref{eq:intercurrent}), we
calculate the interband current by numerically solving the time dependent
Dirac equation (\ref{eq:Diracoriginal}). For comparison, we also calculate the
current for the pseudospin-1/2 system both numerically and analytically. The
results are shown in Fig.~\ref{fig:Interband_J_tE}. For the numerical
results in Fig.~\ref{fig:Interband_J_tE}(a), the momentum space is defined
as $\tilde{p}_x \in [-8, 8]$ and $\tilde{p}_y \in [-8, 8]$ and the integration
grid has the spacing $0.0002$. In Fig.~\ref{fig:Interband_J_tE}(b), we use
the same momentum space grid for $\tilde{E}=0.0001,0.0002,0.0004$ but for
$\tilde{E}= 0.0008, 0.0016, 0.0032$, the ranges of the momentum space are
doubled. From Fig.~\ref{fig:Interband_J_tE}(a), we see that the interband
current for both pseudospin-1 and pseudospin-1/2 cases are independent of
time. That is, after a short transient, the interband current approaches a
constant. From Fig.~\ref{fig:Interband_J_tE}(b), we see that the current
is proportional to the electric field $E$ for both pseudospin-1 and
pseudospin-1/2 particles (with unity slope on a double logarithmic
scale), but the proportional constant is larger in the
pseudospin-1 case. While in the weak field regime, the scaling
relation between the interband current and the electric field is the same
for pseudospin-1 and pseudospin-1/2 particles, there is a striking
difference in the current magnitude. In particular, the interband current
for the pseudospin-1 system is about twice that for the pseudospin-1/2
counterpart, as revealed by both the theoretical approximation
Eq.~(\ref{eq:intercurrent}) and the numerical result [corresponding to
the dashed and solid lines in Fig.~\ref{fig:Interband_J_tE}(a), respectively].
The interband current in the pseudospin-1 system is thus greatly enhanced
as compared with that in the pseudospin-1/2 system.

\begin{figure}
\centering
\includegraphics[width=\linewidth]{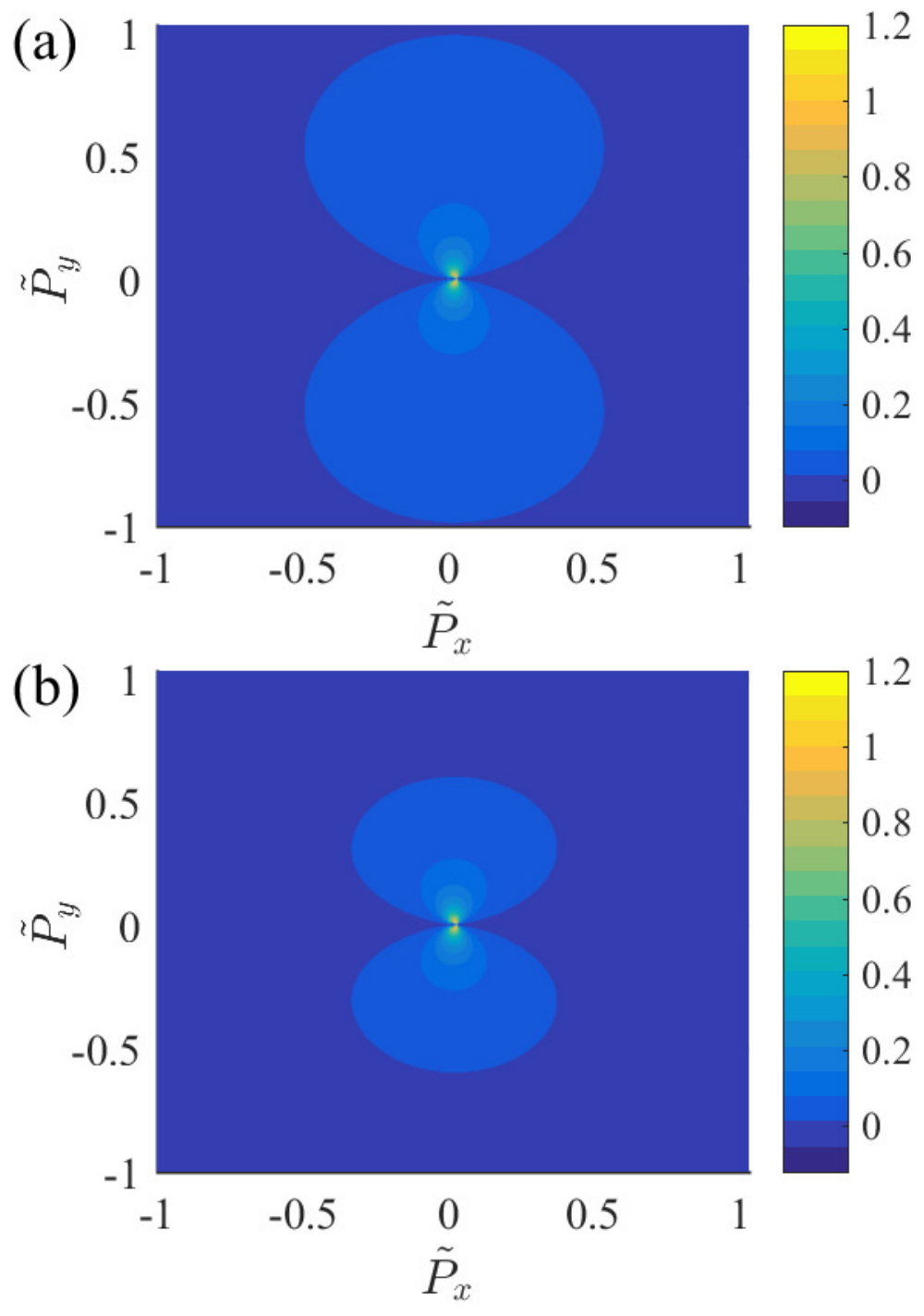}
\caption{ {\bf Interband current distribution in the momentum space}:
(a) pseudospin-1 and (b) pseudospin-1/2 systems. The time and
electric field strength are $\tilde{t}=2$ and $\tilde{E}=0.0128$
respectively.}
\label{fig:Interband_dist1}
\end{figure}

Intuitively, the phenomenon of current enhancement can be
attributed to the extra flat band in the pseudospin-1 system: while
the band itself does not carry any current, it can contribute to the
interband current. Indeed, the theoretical results in
Eqs.~(\ref{eq:interpositive}) and (\ref{eq:interflat}) indicate that
the flat band contributes to the total interband current, while the
positive band contributes little to the current. To gain
physical insights, we numerically calculate three currents: the positive
and flat band currents from the pseudospin-1 system, and the current
from the pseudospin-1/2 system. Figure~\ref{fig:Interband_flat} shows
that the ratio of the flat band current to the pseudospin-1/2 current is
two, while the ratio between the positive band and pseudospin-1/2 currents
is nearly zero, indicating that in the pseudospin-1 system, almost all the
interband current originates from the flat band.


To better understand the phenomenon of interband current enhancement in
the pseudospin-1 system, we calculate the current distribution for both
pseudospin-1 and pseudospin-1/2 systems in the momentum space, as shown
in Fig.~\ref{fig:Interband_dist1}. We see that the area in the momentum
space with significant current is larger for the pseudospin-1 case,
although the current magnitude is almost the same near the Dirac point
for both systems. This is indication that the flat band can contribute
substantially more current because the Landau-Zener transition ``gap''
$P_y$ for the pseudospin-1 system is small compared to that for the
pseudospin-1/2 system. Mathematically, with respect to the single state
current expression (\ref{eq:interflat_approx}) for the pseudospin-1 system,
the corresponding one state contribution to the current for the 
pseudospin-1/2 system is
\begin{equation}
J_p^{half} \approx  \frac{e^2 \hbar E}{2} \frac{\sin^2 \theta}{p^2}\sin{(\frac{2v_F pt}{\hbar})}.
\end{equation}
The integration of current over the entire momentum space gives the
factor 2 of enhancement for the pseudospin-1 system as compared with
the pseudospin-1/2 system. This implies that quantum interference occurs
mainly between particles from the negative and flat bands due to the
small gap between them.

\begin{figure}[!htp]
\centering
\includegraphics[width=\linewidth]{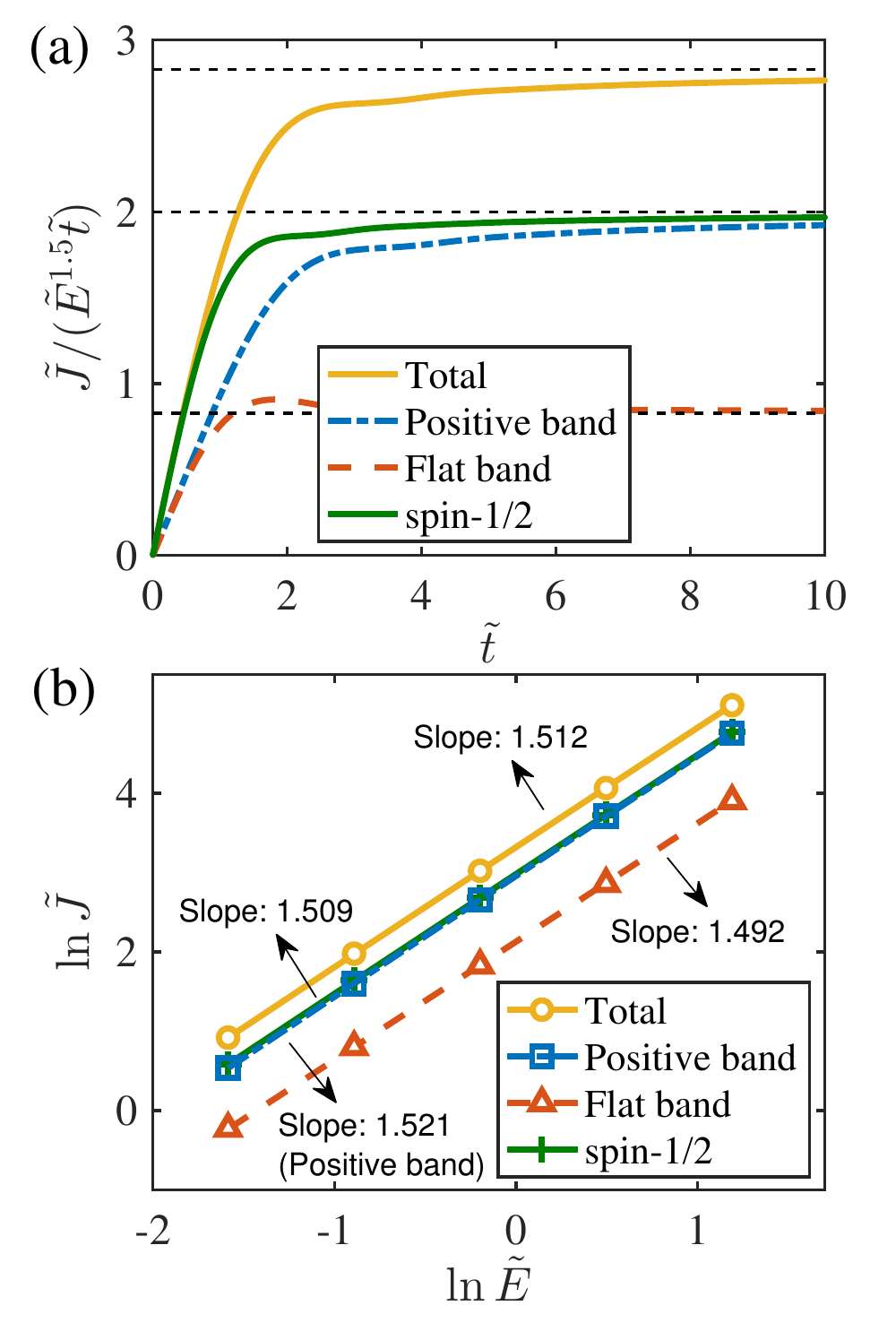}
\caption{ {\bf Enhancement of intraband current in the strong electric
field regime}. Intraband current and contributions from distinct
bands (a) versus time for $\tilde{E}=0.8192$, where the black dashed lines
represent the analytical values $2(\sqrt{2} -1)$, $2$, $2\sqrt{2}$ (from
bottom) and (b) versus electric field at time $\tilde{t}=10$ (for six
values of the electric field: $\tilde{E} = 0.2048, 0.4096, 0.8192, 1.6384,
3.2768$.}
\label{fig:Intraband_J_tE}
\end{figure}

\section{Strong field regime: enhancement of intraband current}
\label{sec:strong_field}

In the strong field regime, 
the intraband current [the first term in 
Eq.~(\ref{eq:current2})] dominates (see Appendix B).
The transition probabilities for the positive, flat and negative bands 
are given, respectively, by~\cite{carroll1986generalisation}
\begin{align}
n_p^{+} &= \Theta (p_x) \Theta (eEt - p_x) \exp (-\frac{\pi v_F p_y^2}{\hbar e E}), \label{eq:positive}\\
n_p^{0} &= \Theta (p_x) \Theta (eEt - p_x) \nonumber \\
&\cdot 2 \Big[ 1 - \exp (-\frac{\pi v_F p_y^2}{2\hbar e E}) \Big] \Big[ \exp (-\frac{\pi v_F p_y^2}{2\hbar e E}) \Big], \label{eq:flaten}\\
n_p^{-} &= \Theta (p_x) \Theta (eEt - p_x)\Big[ 1 - \exp (-\frac{\pi v_F p_y^2}{2\hbar e E}) \Big]^2, \label{eq:negative}
\end{align}
subject to the momentum constraint: $(p_x, eEt - p_x) \gg |p_y|$. The
transition probabilities are essentially the pair production or transition
probabilities in the generalized three-level Landau-Zener model. Substituting
Eqs.~(\ref{eq:positive}) and (\ref{eq:negative}) into Eq.~(\ref{eq:current1}) 
[or equivalently Eq.~(\ref{eq:current2})] and integrating its first term over 
the momentum space, 
we obtain the positive-band contribution to the intraband current with
conducting electrons (or partially filled electrons) populated from the
filled bands
\begin{align}
J^{+}&= \frac{ev_F}{\hbar^2 \pi^2}\iint\frac{v_F(eEt -p_x)}{\epsilon_p(t)} \cdot |\alpha_p(t)|^2  dp_x dp_y \nonumber \\
&\approx \frac{ev_F}{\hbar^2 \pi^2} \int_{0}^{eEt}dp_x \int_{-p_x}^{p_x} |\alpha_p(t)|^2 dp_y \nonumber \\
&\approx \frac{ev_F}{\hbar^2 \pi^2} \int_{0}^{eEt}dp_x \int_{-\infty}^{+\infty} |\alpha_p(t)|^2dp_y \nonumber \\
&= \frac{e^2}{\hbar \pi^2}\sqrt{\frac{ev_F}{\hbar}} E^{3/2} t \tag{20}\\
& = \frac{e\Delta^2}{v_F \hbar^2 \pi^2} \tilde{E}^{3/2} \tilde{t}. \tag{21} \label{eq:J_intra_positive}
\end{align}
The contribution to the current from the initially filled negative band with 
holes left by the electrons driven into the positive and flat bands, the 
conducting hole based intraband current $J^{-}$, is given by
\begin{align}
J^{-} &= (2\sqrt{2} - 1)\frac{e^2}{\hbar \pi^2}\sqrt{\frac{ev_F}{\hbar}} E^{3/2} t \tag{22}\\
& = \frac{e\Delta^2}{v_F \hbar^2 \pi^2} (2\sqrt{2}-1)\tilde{E}^{3/2} \tilde{t}, \tag{23}
\end{align}
which can be written as
\begin{align}
J^{-} = J^{-}_{positive} + J^{-}_{flat}, \tag{24}
\end{align}
where the first term accounts for the contribution by the holes left by
electrons finally driven into the positive band only while the second
term represents the current contribution associated with the hole
concentration induced by the flat band. We have $J^{-}_{positive}=J^{+}$.
The flat band induced current results from the hole concentration 
in the dispersive band, which can be written as
\begin{align}
J^{-}_{flat} &=J^{-} - J^{+} \nonumber\\
& = \frac{e\Delta^2}{v_F \hbar^2 \pi^2}2(\sqrt{2}-1)\tilde{E}^{3/2} \tilde{t}. \tag{25}
\end{align}
Taking into account both the conducting electrons and the corresponding
holes, we obtain the following expression for the dispersive positive band
based current:
\begin{align}
J_{positive} &= J^{+}+J^{-}_{positive}=2\cdot \frac{e^2}{\hbar \pi^2}\sqrt{\frac{ev_F}{\hbar}} E^{3/2} t \tag{26}\\
& = 2\cdot \frac{e\Delta^2}{v_F \hbar^2 \pi^2} \tilde{E}^{3/2} \tilde{t}. \tag{27}
\end{align}
Note that, for the pseudospin-$1/2$ system, this is the total current
in the strong field regime. The total intraband current in the presence
of the flat band in the pseudospin-$1$ system is
\begin{align}
J^{intra} &= J^{+} + J^{-} = J_{positive} + J^{-}_{flat} \nonumber \\
&= 2\sqrt{2}\frac{e^2}{\hbar \pi^2}\sqrt{\frac{ev_F}{\hbar}} E^{3/2} t \tag{28}\\
& = \frac{e\Delta^2}{v_F \hbar^2 \pi^2} 2\sqrt{2}\tilde{E}^{3/2} \tilde{t}. \tag{29}
\label{eq:J_intra_total}
\end{align}
Comparing with the pseudospin-$1/2$ case, we see that the current
enhancement is due to the enhanced hole concentration as a result of the
additional flat band.

The intraband current scales with the electrical field as $E^{3/2}$ and
scales linearly with time, which are the same as those for the pseudospin-1/2
system~\cite{dora2010nonlinear}. However, for the pseudospin-1 system, the
magnitude of the intraband current is larger: there is an enhancement
factor of $\sqrt{2}$ as compared with the pseudospin-1/2 system.
Since the positive band contribution is the same as for the pseudospin-1/2
system, the enhancement is due entirely to the flat band contribution.

\begin{figure}
\centering
\includegraphics[width=\linewidth]{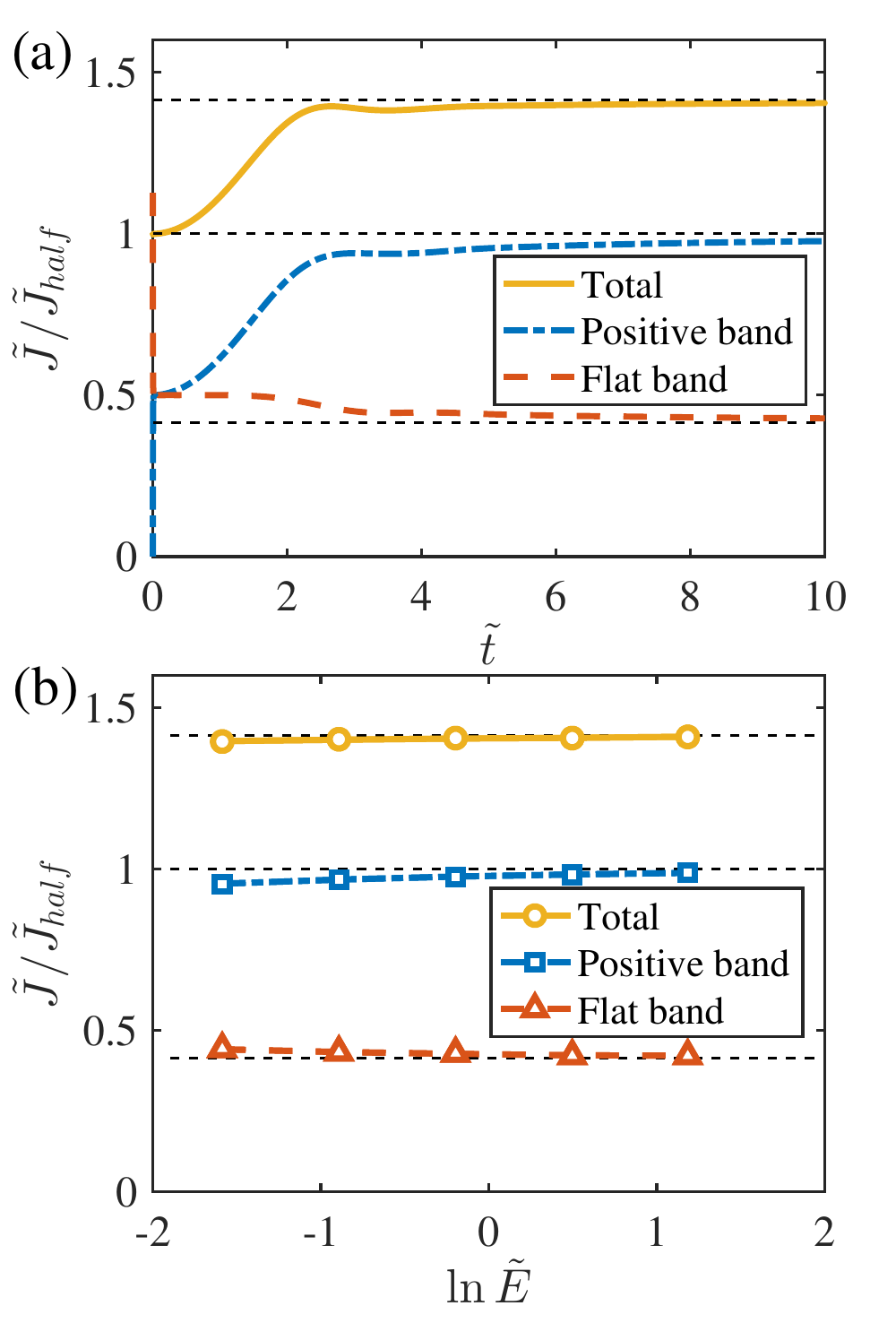}
\caption{ {\bf Further evidence of enhancement of intraband current in
the pseudospin-1 system}. (a) The ratio of the intraband currents in the
pseudospin-1 and pseudospin-1/2 systems versus time $\tilde{t}$ for
$\tilde{E} = 0.8192$. (b) The current ratio versus $\tilde{E}$ for
$\tilde{t} = 10$.}
\label{fig:Intraband_J_ratio_tE}
\end{figure}

We now provide numerical evidence for the predicted phenomenon of intraband 
current enhancement in the pseudospin-1 system. 
Figures~\ref{fig:Intraband_J_tE}(a) and \ref{fig:Intraband_J_tE}(b) show 
the intraband current versus time $\tilde{t}$ and 
the electric field strength $\tilde{E}$, respectively, where the
momentum space grid is $p_x \in [-16, 16]$ and $p_y \in [-16, 16]$ with
spacing $0.002$ in (a) and the momentum space range is increased according to
the increase in the electric field strength in (b). We see that the intraband
current scales with $E$ as $E^{3/2}t$ - the same as for the pseudospin-1/2
system~\cite{dora2010nonlinear,rosenstein2010ballistic}. There is a good
agreement between the numerical results and the theoretical predictions
Eqs.~({\ref{eq:J_intra_positive}}-{\ref{eq:J_intra_total}}).

\begin{figure}
\centering
\includegraphics[width=\linewidth]{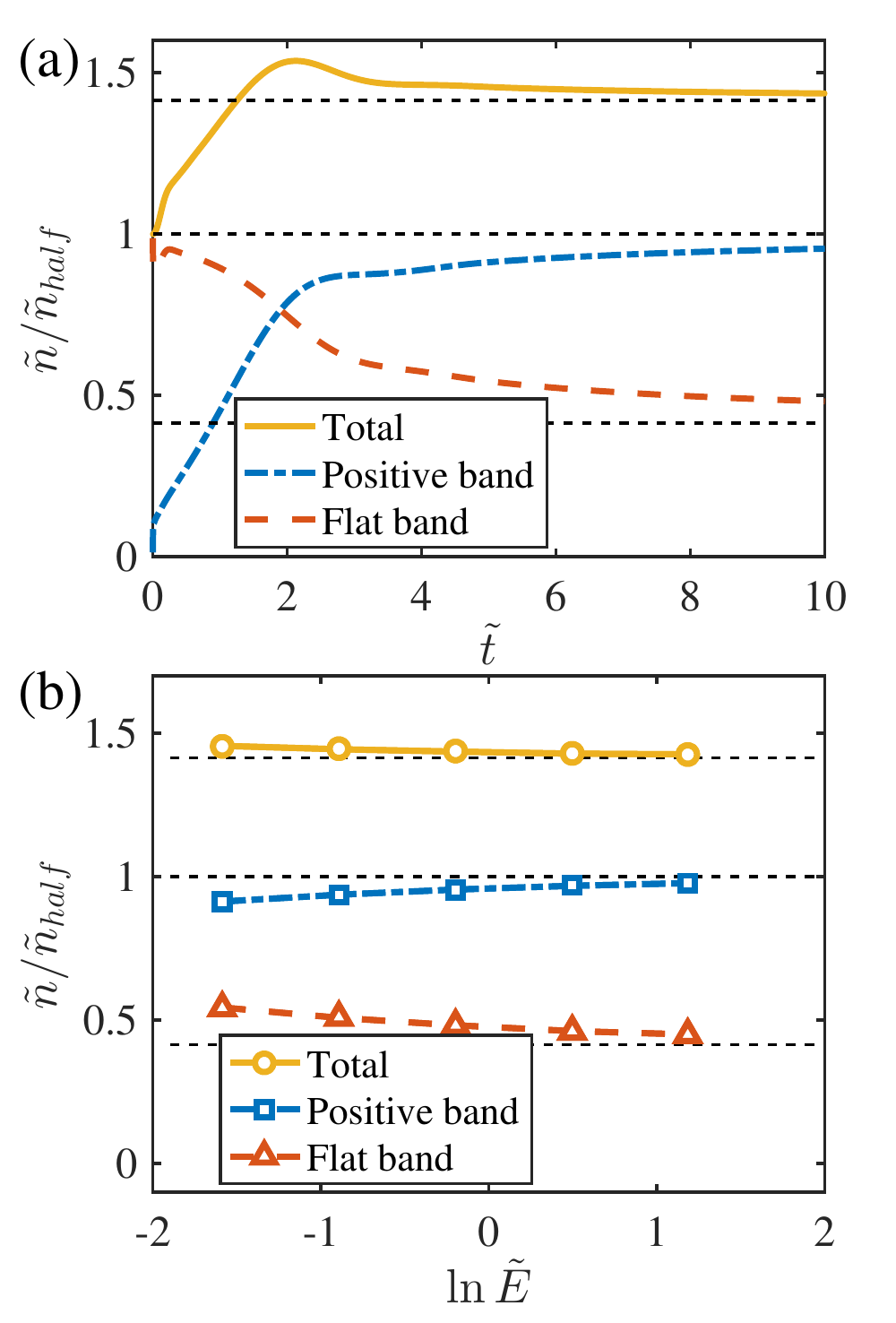}
\caption{ {\bf Numerical evidence of pair creation mechanism for the
intraband current}. The ratio of particle number distribution for
pseudospin-1 and pseudospin-1/2 systems (a) versus time $\tilde{t}$ for
$\tilde{E} = 0.8192$ and (b) versus $\tilde{E}$ for $\tilde{t} = 10$.}
\label{fig:Intraband_n_ratio_tE}
\end{figure}

To provide further confirmation of the enhancement of the intraband current,
we calculate the ratio between the currents from the pseudospin-1 and
pseudospin-1/2 systems versus time for certain electric field, as shown
in Fig.~{\ref{fig:Intraband_J_ratio_tE}}(a). The ratio versus the electric
field for a given time is shown in Fig.~{\ref{fig:Intraband_J_ratio_tE}}(b).
We see that, in the long time regime, under a strong electric field the
total intraband current for the pseudospin-1 system is about $\sqrt{2}$
times the current of the pseudospin-1/2 system. However, the positive band
currents are approximately the same for both systems. The extra current
in the pseudospin-1 system, which is about 0.4 times the contribution
from the positive band, is originated from the flat band. These numerical
results agree well with the theoretical predictions. The physical mechanism
underlying the intraband current enhancement is the Schwinger mechanism or
Landau-Zener tunneling. Note that, in Fig.~{\ref{fig:Intraband_J_ratio_tE}},
the transition of an electron from the negative to the flat bands does not
contribute to the intraband current, as the process leaves behind a hole
in the negative band that contributes to the net current.

\begin{figure}
\centering
\includegraphics[width=\linewidth]{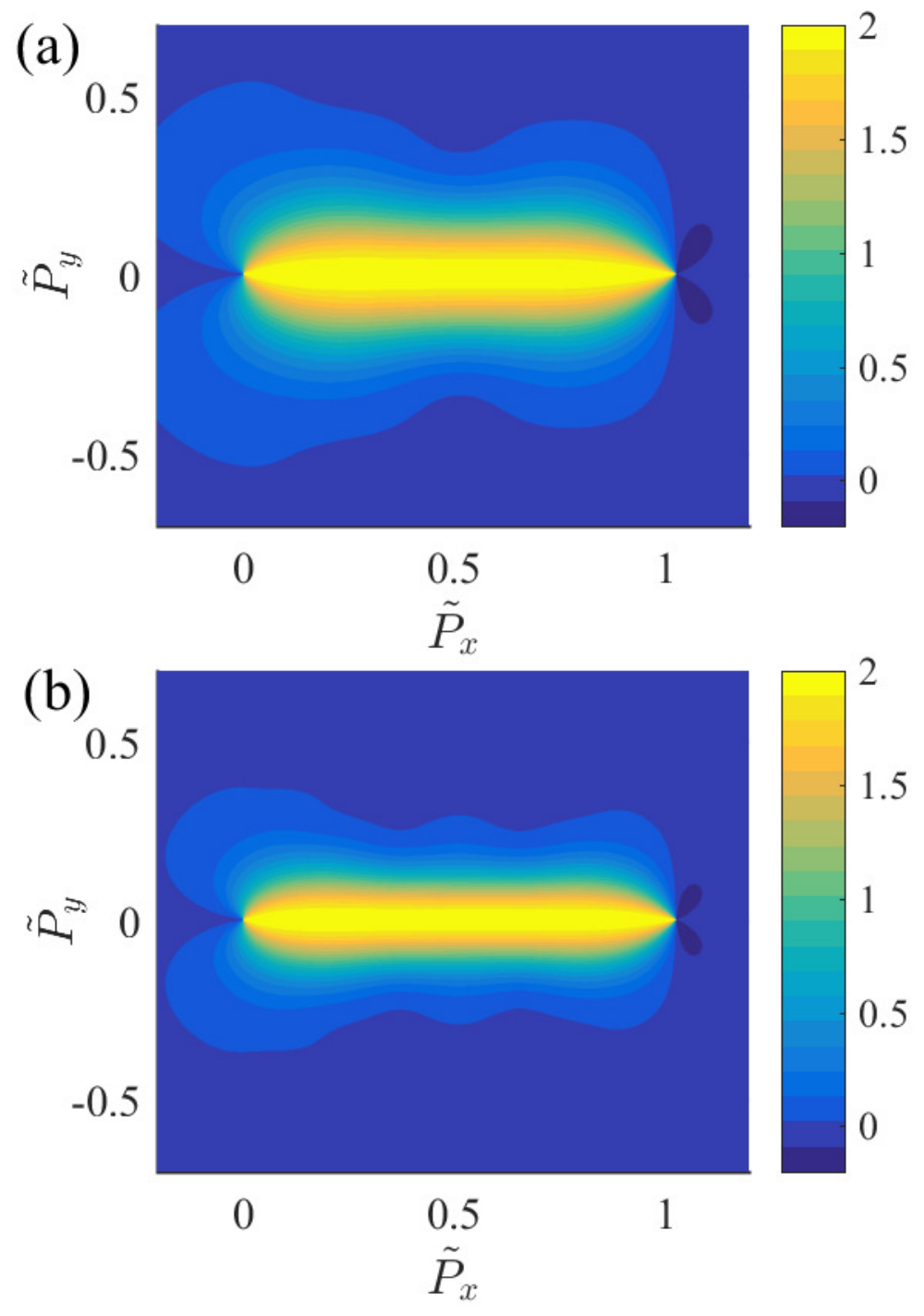}
\caption{ {\bf Current density distribution in the momentum space}.
(a,b) For pseudospin-1 and pseudospin-1/2 systems, respectively, the
distributions of the current density in the momentum space for
$\tilde{t}=20$ and $\tilde{E}=0.0512$. When the momentum gap value
$P_y$ is large, the flat band can enhance the current.}
\label{fig:Intraband_dist}
\end{figure}

If the intraband current is generated by pair creation through Landau-Zener
tunneling, the number of created particles should be consistent with
the current behaviors. To test this, we numerically calculate the particle
number distribution in different bands and plot the ratio between the numbers
of particles for pseudospin-1 and pseudospin-1/2 systems versus time and the
electric field, as shown in Fig.~\ref{fig:Intraband_n_ratio_tE}. For the
pseudospin-1 system, the number of particles created in the positive band
is approximately the same as that created in the upper band in the
pseudospin-1/2 system, and the number of particles in the flat band is
about half of that in the positive band. Note that, for the positive
band, it is necessary to count the particle number twice as both electrons
and holes contribute to the transport current. However, for the flat band,
only holes contribute to the current. We see that, for each band, the
particle number distribution is consistent with the current distribution,
providing strong evidence that the intraband current
results from pair creation in the negative band. In fact, under the
strong field approximation, the intraband current is the particle
distributions in the positive and flat bands multiplying by the constant
$ev_F$, as current is due to electron and hole transport.

We also calculate the current density distribution in the momentum space
for a fixed time and electric field strength, as shown in
Fig.~\ref{fig:Intraband_dist}. We see that the current distribution range
in the $P_y$ direction is wider for the pseudospin-1 system than for the
pseudospin-1/2 system. However, the current distribution near $P_y=0$ is
approximately the same for the two systems, and the current decays in the
$p_y$ direction. In addition, there is a current cut-off about
$\tilde{p}_x = \tilde{E}\tilde{t}$ along the $p_x$ axis. All these features
of the current density distribution can be fully explained by the
theoretical formulas (\ref{eq:positive}-\ref{eq:negative}). The general
result is that the flat band can enhance the current when the ``gap''
$P_y$ is large.

\vspace*{0.2in}

\section{Conclusion and Discussion} \label{sec:conclusion}

We investigate non-equilibrium transport of quasiparticles subject to
an external electric field in the pseudospin-1 system arising from solid
state materials whose energy band structure constitutes a pair of Dirac
cones and a flat band through the conical connecting point. Since the group
velocity for carriers associated with the flat band is zero, one may
naively think that the flat band would have no contribution to the current.
However, we find that the current in the pseudospin-1 system is generally
enhanced as compared with that in the counterpart (pseudospin-1/2) system. In
particular, in the weak field regime, for both systems the interband
current dominates, is proportional to the electric field strength, and
is independent of time. However, the interference between quasiparticles
associated with the flat and the negative bands in the pseudospin-1
system leads to an interband current whose magnitude is twice the current
in the pseudospin-1/2 system. In the strong field regime, for both 
systems the intraband current dominates and scales with the electric
field strength as $E^{3/2}$ and linearly with time. We find that the
current associated with carrier transition from the negative to the
positive bands is identical for both systems, but the flat band
in the pseudospin-1 system contributes an additional term to the current,
leading to an enhancement of the total intraband current. The general
conclusion is that, from the standpoint of generating large current, the
presence of the flat band in the pseudospin-1 system can be quite beneficial.
Indeed, the interplay between the flat band and the Dirac cones can lead to
interesting physics that has just begun to be understood and exploited.

We discuss a few pertinent issues.

\paragraph*{\bf Time scale of validity of effective Dirac Hamiltonian.}
For a real material, the effective Dirac Hamiltonian 
description is valid about the degeneracy (Dirac) point only, imposing 
an intrinsic upper bound on time in its applicability. Similar to the 
situation of using the two-band Dirac Hamiltonian to describe 
graphene~\cite{rosenstein2010ballistic}, such a time bound can be 
approximately estimated as the Bloch oscillation period, i.e., the time 
required for the electric field to shift the momentum across the Brillouin 
zone: $\Delta p_x = eEt \approx \hbar/a$ with $a$ being the lattice constant. 
We obtain $t_B \sim \hbar/(eEa)$. Since the aim of our work is to investigate 
the physics near the Dirac point, the effective Hamiltonian description is 
sufficient. For clarity and convenience, all the calculations are done in terms 
of dimensionless quantities through the introduction of an auxiliary energy 
scale $\Delta$ whose value can be properly set to make the calculations 
under the restriction relevant to the real materials hosting pseudospin-$1$ 
quasiparticles. More specifically, the estimated time restriction $t<t_B$ 
gives rise to the following condition in terms of the dimensionless quantities
$$\tilde{E}\tilde{t}<\frac{\hbar v_F}{\Delta a}.$$
For the given values of $\tilde{t}$ and the range of $\tilde{E}$ in all
figures, the condition is fulfilled by setting $\Delta = \hbar v_F/50a$,
based on which the actual physical units can be assigned to the dimensionless
quantities. It is possible to test the results of this paper experimentally 
through tuning the characteristic energy $\Delta$ of the underlying system.
While our work uses a model Hamiltonian to probe into the essential physics 
of pseudospin-1 systems in a relatively rigorous manner, the issue of 
dissipation (in momentum or energy) is beyond the intended scope 
of this paper.

\paragraph*{\bf Bloch oscillations.}
If the whole band structure is taken into account, Bloch oscillations will
occur under an external electric field for $t\gtrsim t_B$, i.e., the
electron distribution will oscillate over a certain range of the lattice sites.
In this case, the Dirac Hamiltonian description will no longer be valid.
Instead, a full tight-binding Hamiltonian $H_{TB}(\boldsymbol{p})$
characterizing the multiband structure associated with a particular lattice
configuration should be used. For the dice or $T_3$ lattice with intersite
distance $a$ and hopping integral $t$, the tight-binding Hamiltonian is
\begin{eqnarray}
\nonumber
& & H_{TB}^{(dice)}(\boldsymbol{p}) = \begin{bmatrix}
0 & h_{\boldsymbol{p}} & 0 \\
h_{\boldsymbol{p}}^* & 0 & h_{\boldsymbol{p}} \\
0 & h_{\boldsymbol{p}}^* & 0
\end{bmatrix}, \\ \nonumber
& & h_{\boldsymbol{p}} = -t\left(1 + 2\exp{(3ip_ya/2)}\cos(\sqrt{3}p_xa/2)\right).
\end{eqnarray}
A previous work~\cite{rosenstein2010ballistic} showed that, for the honeycomb 
lattice, the corresponding two-band tight-binding model can indeed give rise 
to Bloch oscillations for $t > t_B$. To investigate Bloch oscillations
in the large time regime for pseudospin-1 systems with an extra flat
band is certainly an interesting issue that warrants further efforts.

We note that, in a recent paper~\cite{KF2016}, the striking phenomenon of 
tunable Bloch oscillations was reported for a quasi one-dimensional diamond 
lattice system with a flat band under perturbation. It would be interesting to
extend this work to two-dimensional lattices. The main purpose of our work is 
to uncover new phenomena in physical situations where the Dirac Hamiltonian 
description is valid (first order expansion of the tight binding Hamiltonian 
about the Dirac points).

\paragraph*{\bf Effect of band anisotropy.}
For a particular lattice configuration associated with a real material,
band anisotropy, e.g., the trigonal warping, will generally arise when
entering the energy range relatively far from the Dirac points at
a later time. In this case, direction dependent transport behavior can
arise. Insights into the phenomena of driving direction resolved Bloch
oscillations and Zener tunneling can be gained from existing studies of
the two-band systems with the so-called ``semi-Dirac'' spectrum (a hybrid
of the linear and quadratic dispersion)~\cite{lim2012bloch,lim2014mass}. 
At the present, the interplay between an additional flat band and dispersion 
anisotropy remains largely unknown, which is beyond the applicable scope 
of the idealized Dirac Hamiltonian framework.

\section*{Acknowledgement}

We thank Dr. Guang-Lei Wang for helpful discussions, and would like to
acknowledge support from the Vannevar Bush Faculty Fellowship program
sponsored by the Basic Research Office of the Assistant Secretary of
Defense for Research and Engineering and funded by the Office of Naval
Research through Grant No.~N00014-16-1-2828. L.H. was supported by NSF
of China under Grant No.~11422541.

\appendix
\section{Analytic calculation of the interband current}

In the weak field regime, we can expand Eq.~(\ref{eq:Diracoriginal}) as
\begin{align}
i\hbar \partial_t\alpha_p(t) &= \epsilon_p \alpha_p(t) + C_0 \gamma_p(t), \label{eq:alpha} \\
i\hbar \partial_t\gamma_p(t) &= C_0[\alpha_p(t) + \beta_p(t)], \\
i\hbar \partial_t\beta_p(t)  &= -\epsilon_p \beta_p(t) + C_0 \gamma_p(t) \label{eq:beta}.
\end{align}
Applying the time differential operator $i\hbar \partial_t$ to
Eqs.~(\ref{eq:alpha}) and (\ref{eq:beta}), we get
\begin{align}
i\hbar \partial_t (i\hbar \partial_t \alpha_p(t)) = \epsilon_p i\hbar \partial_t\alpha_p(t) + C_0 i\hbar \partial_t\gamma_p(t), \label{eq:alpha2}\\
i\hbar \partial_t (i\hbar \partial_t \beta_p(t)) = -\epsilon_p i\hbar \partial_t\beta_p(t) + C_0 i\hbar \partial_t\gamma_p(t), \label{eq:beta2}
\end{align}
and, hence,
\begin{equation} \label{eq:couple1}
-\hbar^2 \partial_t^{2}\alpha_p(t)-\hbar^2 \partial_t^{2}\beta_p(t) = [\alpha_p(t) + \beta_p(t)][\epsilon_p^{2} + 2C_0^{2}].
\end{equation}
From Eqs.~(\ref{eq:alpha}) and (\ref{eq:beta}), we have
\begin{equation} \label{eq:couple2}
i\hbar\partial_t\alpha_p(t) - i\hbar\partial_t\beta_p(t)=\epsilon_p [\alpha_p(t) + \beta_p(t)].
\end{equation}
Defining $x_p(t)=\alpha_p(t)+\beta_p(t)$, and $y_p(t)=\alpha_p(t)-\beta_p(t)$,
we get, from Eqs.~(\ref{eq:couple1}) and (\ref{eq:couple2}), respectively,
the following relations:
\begin{align}
&\frac{d^2x_p}{dt^2}+ \frac{\epsilon_p^2 + 2C_0^2}{\hbar^2}x_p = 0, \label{eq:xequation}\\
&\frac{dy_p}{dt} = \frac{\epsilon_p}{i\hbar}x_p. \label{eq:yequation}
\end{align}
Solving Eq.~(\ref{eq:xequation}), we get
\begin{equation}
x_p(t) = A\cos\omega t + B\sin\omega t, \nonumber
\end{equation}
where $A$ and $B$ are constant, and
$\omega = \sqrt{(\epsilon_p^2 + 2C_0^2)/\hbar^2}$. Using the initial
condition that the negative band is fully filled:
($\Phi_p(t=0) = [0, 0, 1]^T$), we have $x_p(t=0) = A = 1$. From
Eq.~(\ref{eq:yequation}), we have
\begin{equation}
y_p(t) = \frac{\epsilon_p}{i\hbar \omega} [\sin\omega t - B\cos\omega t] +d.\nonumber
\end{equation}
Using the initial condition, we get $y_p(t=0) = -m_0B + d = -1$, where
$m_0 = \epsilon_p/(i\hbar \omega)$, $d = m_0 B -1$, which leads to
\begin{align}
&\alpha_p(t) = \frac{1}{2}(x + y) \nonumber \\
&=\frac{1}{2}[\cos\omega t + B \sin\omega t + m_0(\sin\omega t - B\cos\omega t + B) - 1], \nonumber\\
&\beta_p(t) = \frac{1}{2}(x - y) \nonumber \\
&=\frac{1}{2}[\cos\omega t + B \sin\omega t - m_0(\sin\omega t - B\cos\omega t + B) + 1].\nonumber
\end{align}
Substituting the expressions of $\alpha_p(t)$ and $\beta_p(t)$ into
Eqs.~(\ref{eq:alpha}) and (\ref{eq:beta}), we obtain an expression for
$\gamma_p(t)$. Using $\gamma_p(t=0) = 0$, we have $B = -m_0$ and, hence,
\begin{align}
&\alpha_p(t) = \frac{1}{2} [\cos \omega t + m_0^2(\cos \omega t - 1) - 1], \\
&\beta_p(t) = \frac{1}{2} [\cos \omega t - 2m_0 \sin\omega t - m_0^2 [\cos\omega t - 1] + 1], \\
&\gamma_p(t) = \frac{1 + m_0^2}{2C_0} [-i\hbar \omega \sin \omega t - \epsilon_p(\cos \omega t - 1)].
\end{align}

\section{Dominant current source in the weak and strong field regimes}

\begin{figure}[h]
\centering
\includegraphics[width=\linewidth]{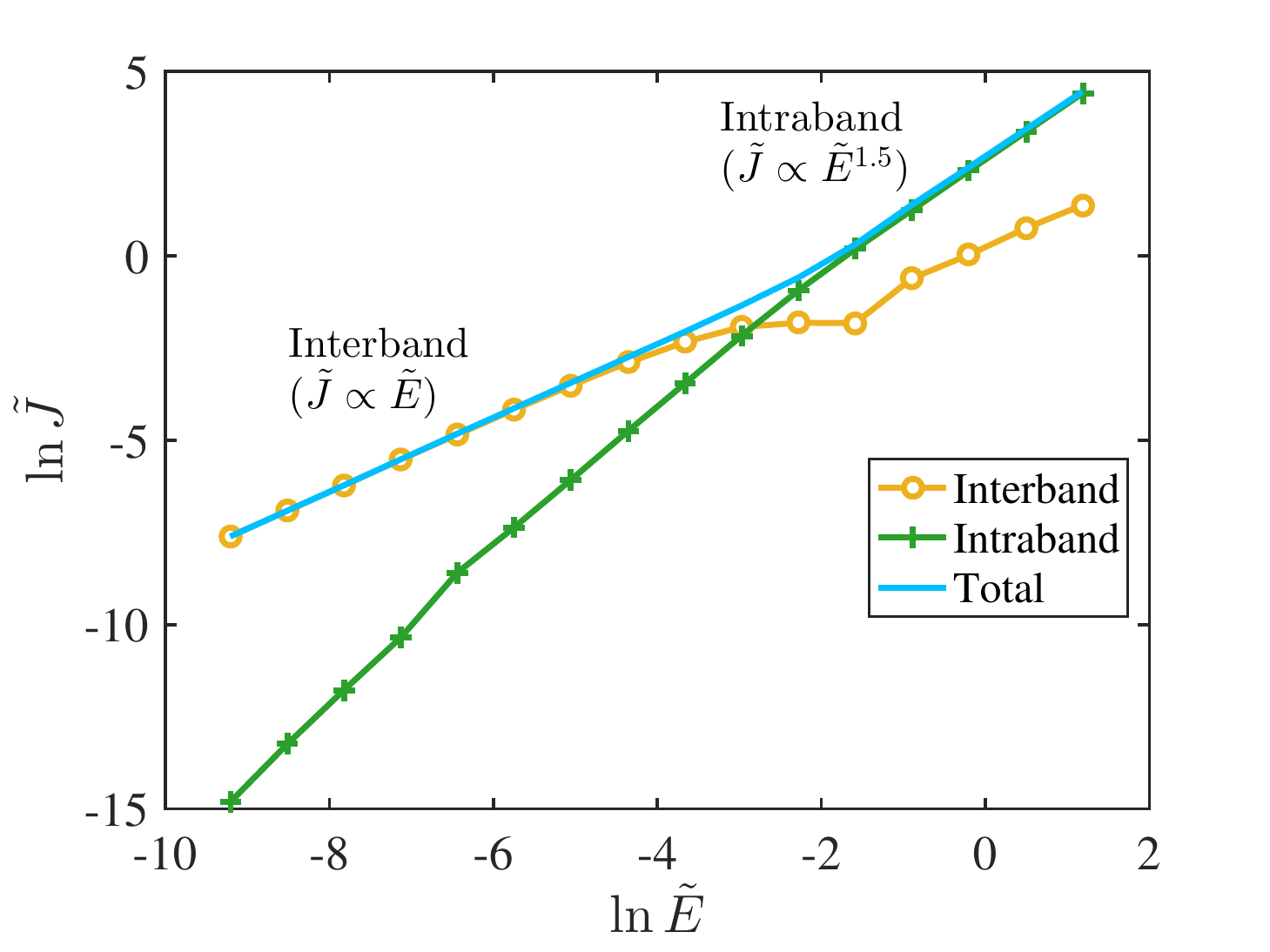}
\caption{ {\bf Current versus electric field of pseudospin-1 system for
$\boldsymbol{\tilde{t} = 5}$}. As the magnitude of the external electrical
field is increased, the dominant contribution to the total current changes
from interband to intraband, and the algebraic scaling exponent of the 
current-field relation changes from 1 to 1.5.}
\label{fig:Inter_intra_dist}
\end{figure}

For the three-band dispersion profile investigated in this work, there
are two distinct current sources: the intraband
and interband currents, where the former is proportional to the number
of electrons (holes) within an unfilled (occupied) band while the latter
depends on the rate of change in the particle number - a characteristic of 
interband interference. From Eq.~(\ref{eq:current2}), we see that the
intraband current is determined by the transition amplitudes while the
interband current depends on the rate of change of the amplitudes. For a weak 
driving field, the transition amplitudes between the occupied and the empty 
bands are negligibly small, so is the number of electron-hole generation, 
resulting in a weak intraband current. However, the rate of change in the 
transition amplitudes may not be small, neither is the interband current. Our 
calculations reveal that, indeed, in the weak (strong) driving regime, the
interband (intraband) current dominates. As the field is increased from the 
weak to the strong regime, the algebraic scaling exponent of the current-field 
relation changes from 1 to 1.5, as shown in Fig.~\ref{fig:Inter_intra_dist}.

%
\end{document}